\documentclass[12pt]{article}
\title{A thought experiment with clocks in static gravity}
\author{L.B.Okun \\
ITEP, Moscow, 117218, Russia \\
Email: okun@heron.itep.ru}
\date{}
\begin{document}
\maketitle
\begin{abstract}
In order to directly demonstrate that in static gravitational field
the rate of clocks increases with their distance from the source a
simple thought experiment is proposed.  \end{abstract}

\section{Introduction}

The redshift of the frequentcy of photons as they travel upwards in
a static gravitational  field and corresponding increase of the
frequency of clocks (e.g. of the spacings between atomic levels) were
theoretically discovered by Einstein long before he constructed
general relativity. These phenomena are discussed in a vast
literature going from monographs on general relativity to popular
books. In the recent articles \cite{1,2,3} it has been argued that in
a static field for a static observer the frequency of the photon does
not change and that it is redshifted only with respect to the
blueshifted atomic (nuclear) levels. Strangely enough, this statement
met with objections from some of the experts in general relativity.
These requested from the authors a thought experiment involving only
clocks but no photons, which could prove that standard clocks are
running faster when they are  raised above the ground. The present
note proposes such a thought experiment. In order to make the {\it
pro} and {\it contra} arguments clearer the following text is written
as a dialogue between a professional relativist (R) and the author
(A).

\section{Dialogue on the thought experiment}

{\it A}: In refs. \cite{1} - \cite{3} two kinds of arguments were
presented: theoretical and experimental. According to the theoretical
argument, the wave solution of Maxwell's equations in a static
gravitational field must have a fixed frequency equal to that of the
e.m. source since the coefficients of Maxwell's equations do in that
case not depend on time.

{\it R}: The conservation of energy (or frequency) of the photon
depends  on the definition, depends on what you choose to call the
energy or frequency, on the definition of time and the system of
reference. To be operationally meaningful you have to specify the way
of measuring these entities.

{\it A}: As we consider a static gravitational field we should choose
observers which are at rest in that field, say in laboratories
situated at ground floor and the upper floor of the same building.
That was how the first measurements of gravitational redshift were
performed in the famous experiments of Pound and Rebka forty years
ago. We should not consider observers in moving elevators. To them
the gravitational field is not static. Moreover in a falling elevator
the gravitational field vanishes locally.

{\it R}: OK. Let choose the static observers and consider the
experiments  by Pound et al. In these experiments photons were
emitted by excited iron nuclei at the ground floor and their
absorption by iron nuclei in their ground state was measured
at the top of Harvard tower. It was observed that the
energy of photons was not sufficient to excite in the iron nuclei
upstairs the same level from which they were emitted downstairs. The
relative shift of the frequency of photons and nuclei was $-gh/c^2$,
where $h$ is the height of the tower, $g=10$m/s$^2$ and $c$ -- the
velocity of light.  One can interpret the experiment by saying (as
the authors of \cite{1} - \cite{3} did) that the energy of the photon
has not changed, but that the energy of excitation of iron nuclei
(the rate of the clock) has fractionally increased by $+gh/c^2$.

Alternatively, there is an equally valid viewpoint according to which
the clock rates are the same while the photon frequency is different.
For that it is enough to consider local inertial frames momentarily
at rest at the world-points of the emission and absorption of the
photon.

{\it A}: But neither Pound, no his coworkers (thank God!) used in
their  experiments falling elevators.

{\it R}:One can introduce local time by simply assuming that standard
clocks do not change their rate with their height. 
Anyway, you are attempting improperly to subdivide the
gravitational  redshift phenomenon into two separate processes, that
separately have no operational meaning. The only operationally
meaningful statement is that when a photon is emitted at the bottom
of the tower its frequency as measured by the identical clock at the
top of the tower is lower. You cannot determine whether it was the
photon or the clocks that were responsible for the shift.

My view would be different if you could carry out the following
thought  experiment. Two identical clocks are placed one at the top
of a tower, the other at the bottom. Demonstrate without connecting
the clocks by photons and otherwise disturbing the clocks that the
top clock runs faster than the bottom clock by the standard
gravitational shift. This would be a demonstration of your assertion
that the photon frequency does not change. I believe such a
demonstration to be impossible.

{\it A}: There were special experiments described in refs.
\cite{1}, \cite{2} in which one of two clocks travelled on an aircraft
and then was brought back to another one which was all this time at
rest in the laboratory. These experiments showed that in
agreement with general relativity the clock which travelled (after
taking account of special relativity twin effect) was ahead of that
which was at rest.

{\it R}: These results do not change my point of view, because the
two  clocks, in the course of the experiment, were treated
differently: one made a loop in space, while the other was not moved
at all. Thus, those experiments demonstrate actually only that the
total elapsed times differ and say nothing (except by inference)
about the instantaneous difference in their rates.

{\it A}: Inference is an inalienable part of physics. But I shall
not dwell on this here. I shall just propose that thought experiment
about which you asserted ``I believe such a demonstration to be
impossible".

Consider two rooms: one at the ground floor, the other at the top
floor of a building. One starts with two identical standard clocks at
the bottom room. At a certain moment one raises the first clock to the
room upstairs. After some time $T$ one raises the second clock in the
same way and compares the time it shows with that of the first clock.
If general relativity is correct, their time difference should be
$\Delta T = (gh/c^2)T$. In this thought experiment both clocks are
treated absolutely identically. Nevertheless the first clock will be
ahead of the second. As time is uniform and both clocks are not moved
during the time $T$, this thought experiment yields the instantoneous
difference in the rates of the clocks, which is of course equal to
$\Delta T/T =gh/c^2$.

\section{Conclusion}

By a thought experiment (which with improvements in the precision of
atomic clocks could one day become a real one) it is shown that when
measuring the ``gravitational redshift of photons" one actually
measures the gravitational blueshift of clocks.

\vspace{5mm}

{\large\bf Acknowledgments}

I am grateful to Prof. R.H.Romer for stimulating me to invent the
thought  experiment described in this note. I am grateful to Val
Telegdi for helpful remarks. This work was supported by RFBR (grant
\# 00-15-96562), by A. Von Humboldt Foundation, and by CERN Theory
Division.

\end{document}